\documentstyle[12pt]{article}
\def\hybrid{\topmargin -20pt    \oddsidemargin 0pt
        \headheight 0pt \headsep 0pt
        \textwidth 6.25in       
        \textheight 9.5in       
        \marginparwidth .875in
        \parskip 5pt plus 1pt   \jot = 1.5ex}

\hybrid

\newcommand{\cD}{{\cal D}}

\newcommand{\cL}{{\cal L}}

\newcommand{\cN}{{\cal N}}

\newcommand{\cR}{{\cal R}}

\newcommand{\ax}{\alpha}
\newcommand{\bx}{\beta}
\newcommand{\cx}{\gamma}

\newcommand{\ab}{\bar\alpha}
\newcommand{\bb}{\bar\beta}
\newcommand{\cb}{\bar\gamma}
\newcommand{\db}{\bar\delta}
\newcommand{\au}{\hat\alpha}
\newcommand{\bu}{\hat\beta}
\newcommand{\cu}{\hat\gamma}
\newcommand{\du}{\hat\delta}
\newcommand{\eu}{\hat\epsilon}
\newcommand{\fu}{\hat\phi}
\newcommand{\vv}{\hat v}
\newcommand{\e}{\sqrt{-g}}
\newcommand{\be}{\begin{equation}}
\newcommand{\ee}{\end{equation}}
\newcommand{\bea}{\begin{eqnarray}}
\newcommand{\eea}{\end{eqnarray}}
\newcommand{\ba}{\begin{array}}
\newcommand{\ea}{\end{array}}
\newcommand{\bi}{\begin{itemize}}
\newcommand{\ei}{\end{itemize}}
\newcommand{\bt}{\begin{tabular}}
\newcommand{\et}{\end{tabular}}
\newcommand{\bc}{\begin{center}}
\newcommand{\ec}{\end{center}}
\newcommand{\SLtwo}{$SL(2,{\bf Z})$}
\newcommand{\half}{\frac12}
\newcommand{\quart}{\frac14}
\def\D{\phi_4}
\def\Da{\tilde \phi}
\def\DA{\phi_{\rm A}}
\def\R{{\rm Re}}
\def\I{{\rm Im}}

\def\Jnote#1{{\bf[JL: #1]}}

\begin {document}
\begin{titlepage}
\begin{center}

\hfill hep-th/9908007\\

\vskip .6in
{\large \bf 
Compactification of Type IIB String Theory
on Calabi-Yau Threefolds
}
\vskip .5in

{\bf Robert B\"ohm$^{a}$, Holger G\"unther$^{a}$, Carl Herrmann$^{b}$, 
 Jan Louis$^{a}$}
\footnote{email:  
robert@hera.physik.uni-halle.de,
guenther@hera.physik.uni-halle.de,
herrmann@cptsu5.univ-mrs.fr,
j.louis@physik.uni-halle.de.}
\\
\vskip 0.8cm
{}$^{a}${\em Martin--Luther--Universit\"at 
Halle--Wittenberg,\\
        FB Physik, D-06099 Halle, Germany}
\vskip 0.5cm
{}$^{b}${\em Centre de Physique Th\'eorique,
CNRS -- Luminy Case 907,\\
F-13288 Marseille Cedex 9, France }

\end{center}

\vskip 2.5cm

\begin{center} {\bf ABSTRACT } \end{center}
We study compactifications of type IIB
supergravity on Calabi-Yau threefolds.
The resulting low energy effective
Lagrangian is displayed in the large
volume limit and its symmetry properties 
-- with specific emphasis on the 
\SLtwo\ -- are discussed.
The explicit map to type IIA string theory
compactified on a mirror Calabi-Yau
is derived.
We argue that strong coupling effects
on the worldsheet break the \SLtwo.

\vfill

August 1999
\end{titlepage}


\section{Introduction}
Compactifications of string theories on 6-dimensional
Calabi-Yau  manifolds result in string vacua 
with four flat Minkowskian dimensions ($d=4$) and, 
depending on the type of Calabi-Yau manifold, 
a fixed number of preserved supercharges. 
Specifically, string vacua with $N=2$ supersymmetry 
in $d=4$
are obtained by
compactifying type II strings on Calabi--Yau
threefolds $Y_3$, the heterotic string on
$K3\times T^2 $ or the type I string 
on $K3\times T^2 $.
It is believed that all of the resulting string
vacua are perturbative descriptions 
of different
regions in one and the same 
moduli space \cite{kv,fhsv}.
In this paper we confine our attention
to the region of the moduli space
where type II string compactifications are 
the appropriate perturbative theories.

In $d=10$ type II string theories come in 
two different versions, the non-chiral type IIA
and the chiral type IIB string theory
but upon compactification on Calabi-Yau threefolds
they are related by mirror symmetry.
More precisely, type IIB compactified on $Y_3$ 
is equivalent to IIA compactified on the mirror
manifold $\tilde Y_3$ \cite{seiberg,cfg}.
Due to this equivalence one has the choice 
to study either compactification
but since there is no simple covariant action
of type IIB supergravity in $d=10$ \cite{schwarz}
type IIA compactifications appear to be easier
\cite{CJ}.
In terms of the underlying conformal field 
theory the two theories in $d=4$ only differ
by the GSO-projection and thus can be discussed
on the same footing \cite{seiberg,cfg}.
However, it is believed
that the type IIB theory in $d=10$
has an exact non-perturbative \SLtwo\
symmetry \cite{HTW} which is not shared 
by type IIA (in $d=10$). 
Thus, it is of interest to study 
the `fate'
of this \SLtwo\ symmetry upon compactification
to $d=4$ \cite{berkovits}
and this is one of the goals of this paper.
For that reason we focus on
geometrical compactifications
(and not the conformal field theory versions)
of the effective type IIB theory for most parts of this paper.
In addition, the general type IIB 
effective Lagrangian and the explicit map
to the type IIA vacua has only 
been given in a special case \cite{BC,IIBref}
and as a byproduct we close this 
gap.

This paper is organized as follows.
In section~2 we recall
some of the properties of type IIB supergravity
in $d=10$. In section~3 we discuss
the compactification on Calabi-Yau threefolds
and display the four-dimensional
low energy effective action in the large volume 
limit.
Apart from the gravitational multiplet 
one finds $h_{(1,1)}$ tensor multiplets,
$h_{(1,2)}$ vector multiplets
and one ``double-tensor'' multiplet.
We concentrate on the couplings
of the tensor multiplets since 
the couplings of the vector multiplets 
have been given before \cite{wp,cfg,BCF}. 
In addition the \SLtwo\ -- 
inherited from the $d=10$ theory -- acts on the 
tensor multiplets while the vector multiplets 
are left invariant. 
In section~4 the explicit map between
type IIB compactified on $Y_3$ and type IIA
compactified on the mirror manifold $\tilde Y_3$
is displayed.
This map enables us to also discuss
the geometry of the tensor multiplets in terms
of a holomorphic prepotential and obtain the 
worldsheet instanton corrections.
We find that the \SLtwo\ is manifest at the string tree
level in the large volume limit of the compactification
but broken once strong coupling effects on the worldsheet
(worldsheet instantons) are taken into account.
Furthermore, the \SLtwo\ is also present
in the large complex structure limit of type IIA
vacua but rather obscure in the standard field variables.
Section~5 contains our conclusions and some
of the technical aspects of this paper 
are collected in an appendix.

\section{Type IIB Supergravity  in $d=10$}
The bosonic massless modes of the type IIB string are
the graviton $g_{MN}\ (M,N = 0, \ldots,9)$, 
a doublet of antisymmetric tensors
$B^I_{MN}\ (I=1,2)$, two real scalar fields $\phi, l$
and a 4-form $\cD_{MNPQ}$ with
a self-dual field strength.
$g_{MN}, B^1_{MN}$ and $\phi$ arise in the NS-NS
sector while $B^2_{MN},l$ and $\cD_{MNPQ}$
reside in the R-R sector;
$\phi$ is the dilaton of type IIB string theory.

The difficulty of constructing
a covariant low energy effective action 
for these massless modes is due to
the presence of the self-dual 5-form field strength.
The field equations of type IIB supergravity
on the other hand are well known 
\cite{schwarz}.
However, if one does not impose the 
self-duality condition 
a covariant action (in the Einstein frame)
can be given \cite{BBO}\footnote{%
Recently a covariant action including
the self-dual
4-form has been constructed
\cite{DLS} but here we choose to
follow the procedure outlined in
refs.~\cite{BBO} for compactifying type IIB 
supergravity and use (\ref{Sten}) instead.}
\bea\label{Sten}
  S &=& -\half \int d^{10}x\, \e\,\Big(\cR - 
\quart Tr(\partial M \partial M^{-1})
+\frac{3}{4} H^I M_{IJ} H^J\nonumber
\\
       &&\quad
+\, \frac{5}{6}\, {{F}}^2\ +
       \frac{1}{96\e}\, {\epsilon}_{IJ}\,
       {\cD}\wedge{{H}}^{I}\wedge{H}^{J}\Big),
\eea     
where $\cR$ is the scalar curvature
 and we abbreviated
\bea\label{Mten}
M_{IJ}&\equiv& \frac{1}{\I \lambda} 
\left(\ba{cc}
|\lambda|^2 & -\R\lambda \\
-\R\lambda & 1
\ea\right),\qquad 
\lambda\equiv l+ie^{-\phi}\ ,\nonumber\\
H^I_{MNP} &\equiv&
     \partial_{[{M}}{{B}}^{I}_{NP]}
= \frac{1}{3}\, (\partial_{M}{{B}}^{I}_{NP} 
+ \partial_{{N}}{{B}}^{I}_{PM}
+ \partial_{{P}}{{B}}^{I}_{MN})\ ,\\ 
F_{MNPQR} &\equiv&
   \partial_{[{M}}\cD_{NPQR]}
   + \frac{3}{4}\, {\epsilon}_{IJ}
     B^I_{[MN}\partial_PB^J_{QR]}\ .\nonumber
\eea
In eq.~(\ref{Sten}) we suppress space-time
indices;
in particular, the topological term 
${\cD}\wedge{{H}}^{I}\wedge{H}^{J}$ is contracted 
with the
ten-dimensional $\epsilon$-tensor while all other terms
are contracted with the ten-dimensional metric. Finally, 
we choose ${\epsilon}_{12}=+1$.
The type IIB supergravity is obtained as the 
Euler-Lagrange equations of the action (\ref{Sten})
together with the self-duality condition
\be\label{selfdual}
F_{MNPQR}=\frac{1}{5!\e}\, {\epsilon}_{MNPQRSTUVW}
  F^{STUVW}\ .
\ee

This theory has a number of symmetries.
First of all there are the gauge transformations
of the 4-form (with parameters $\Sigma_{NPQ}$)
\be\label{Dsymm}
\delta \cD_{MNPQ} = 
\partial_{[M}\Sigma_{NPQ]},
\ee
under which all other fields are invariant.
Secondly, there are the gauge transformations
of the two antisymmetric tensors
(with parameters $\Omega^I_{N}$)
which also transform the 4-form 
\bea\label{symmten}
\delta B_{MN}^I &=& \partial_{[M}\Omega^I_{N]}\ ,
\nonumber\\
\delta \cD_{MNPQ} &=& -\frac34\,
\epsilon_{IJ}\, \Omega^I_{[M}H_{NPQ]}^J\ .
\eea
Finally, there is an \SLtwo\
acting as follows\footnote{%
In fact at the classical level one has the larger
group $SL(2,{\bf R})$ which is broken by quantum
corrections to \SLtwo.}
\bea\label{SL1}
\lambda &\mapsto& 
\lambda'=\frac{a{\lambda}+b}{c{\lambda}+d}\ ,
\nonumber\\
H^I_{MNP} &\mapsto& H^{\prime I}_{MNP} 
= \Lambda^I_J H^J_{MNP}\ ,
\eea
where
\begin{equation}\label{SL2}
\Lambda= \left(\begin{array}{cc}
  d & c \\
  b & a
  \end{array}\right)\ ,\qquad ad-bc=1,
\quad a,b,c,d\in{\bf Z}\ .
\end{equation}
Using (\ref{Mten})
one shows that the matrix $M$ transforms
according to 
\be\label{SL3}
M\mapsto M' = \Lambda^{-1T} M {\Lambda}^{-1}\ .
\ee

\section{Calabi-Yau Compactifications of 
Type IIB Supergravity}
\subsection{The Spectrum}
Let us now study the compactification of the 
type IIB low energy effective theory
on Calabi-Yau threefolds $Y_3$.
The resulting theory in $d=4$ has  $N=2$ supersymmetry
and the low energy spectrum comes in
appropriate $N=2$ supermultiplets.
More specifically one finds that the 
10-dimensional metric $g_{MN}$
decomposes into the 4-dimensional metric
$g_{\mu\nu}\ (\mu,\nu =0,\ldots,3)$, 
$2\times h_{(1,2)}$ deformations of the
complex structure $\delta g_{\ax\bx}, 
\delta g_{\ab\bb}\ 
(\ax,\bx = 1,2,3)$
and $h_{(1,1)}$ deformations of the
K\"ahler class $\delta g_{\ax\bb}$.
The Hodge numbers
$h_{(1,1)} (h_{(1,2)})$ count the harmonic 
$(1,1)$-forms $((1,2)$-forms)
on $Y_3$.
The antisymmetric tensors $B_{MN}^I$
decompose into a doublet of antisymmetric 
tensors  $B_{\mu\nu}^I$ and 
$2\times h_{(1,1)}$ scalar modes $B_{\ax\bb}^I$.
Finally the 4-form
decomposes into
$h_{(1,1)}$ (real) antisymmetric tensors
$\cD_{\mu\nu \ax\bb}$ and $h_{(1,2)}+1$ real 
vectors $\cD_{\mu \ax\bx\cb}, 
\cD_{\mu \ax\bx\cx} $.\footnote{
Note that the dimensionally reduced 
self-duality condition (\ref{selfdual})
relates $\cD_{\ax\bx\ab\bb}$ to 
$\cD_{\mu\nu \cx\cb}$,
$\cD_{\mu \ax\bx\cx}$ to $\cD_{\mu\ab\bb\cb}$
and $\cD_{\mu \ax\bx\cb}$ to $\cD_{\mu\ab\bb\cx}$.}
Together with the two scalars $\phi$ and $l$
these fields assemble into the following
$N=2$ supermultiplets:
$$
\ba{ll}
{\rm gravitational\ multiplet}&
(g_{\mu\nu}, \cD_{\mu\ax\bx\cx })\ ,\nonumber\\
{\rm double-tensor\ multiplet} &
(B_{\mu\nu}^I,\phi,l)\ ,\\
h_{(1,2)}\ {\rm vector\ multiplets} &
(\cD_{\mu \ax\bx\cb}, \delta g_{\ax\bx}, 
\delta g_{\ab\bb})\ ,\\
h_{(1,1)}\ {\rm tensor\ multiplets}&
(\cD_{\mu \nu \ax\bb}, \delta g_{\ax\bb}, B_{\ax\bb}^I)\ ,\nonumber
\ea
$$
where we only display the bosonic components.
The gravitational multiplet contains 2 gravitini
while all the other multiplets contain
two Weyl fermions.\footnote{%
The double-tensor multiplet has not been 
constructed as an off-shell multiplet
but we expect that it exists
since it appears in the low
energy limit of type IIB string theory.} 
In $d=4$ an antisymmetric tensor
describes one physical degree of freedom
and thus can always be dualized to a scalar.
At the level of supermultiplets
this duality relates both the tensor and 
the double-tensor multiplet to a 
hypermultiplet which contains four real
scalar degrees of freedom.
In this dual basis the low energy spectrum
features  $h_{(1,1)}+1$ hypermultiplets and
$h_{(1,2)}$ vector multiplets apart from the gravitational multiplet.

The couplings of the vector multiplets have been
obtained before \cite{wp,cfg,BCF}
and thus we exclusively
focus on the couplings of
the $h_{(1,1)}$ tensor and the universal
double-tensor multiplet. (Or in other words
we consider Calabi-Yau threefolds with $h_{(1,2)}=0$.)
On a Calabi-Yau threefold the 
K\"ahler deformations of the metric can be expanded
in terms of harmonic $(1,1)$-forms $\omega^a_{\ax\bb}$
according to 
\bea\label{harmonic}
\delta g_{\ax\bb} &=& 
\vv^a(x)\ \omega^a_{\ax\bb}\ ,\qquad
B_{\ax\bb}^I\ =\ \hat b^{Ia}(x)\ \omega^a_{\ax\bb}\ ,
\nonumber\\
\cD_{\mu \nu\ax\bb}&=& \hat D^a_{\mu \nu}(x)\ 
\omega^a_{\ax\bb}\ ,
\qquad a=1,\ldots, h_{(1,1)}\ .
\eea
In terms of these variables the tensor multiplets
consist of 
$(\hat D_{\mu \nu}^a, \vv^a, \hat b^{Ia})$.

Before we display the four-dimensional Lagrangian
let us collect a few more properties of
Calabi-Yau threefolds \cite{AS,CdO}. 
The K\"ahler form $J$ 
is defined as
\be
J=i\delta g_{\ax \bb}\, d\xi^\ax\wedge 
d{\bar{\xi}}^{\bb}\ ,
\ee
where the $\xi^\ax$ are the complex coordinates
on $Y_3$.
%
The volume $V$ can be expressed in terms of $J$
according to 
\be
V=
i\int \sqrt{g} d^6\xi = \frac{1}{6}
\int J \wedge J \wedge J \ .
\ee
On the space of $(1,1)$-forms one  defines
a metric 
\be\label{metricdef}
G_{ab}=-\frac{i}{V}\int \omega^a_{\ax\bb}\,
\omega^b_{\cx\db}\,
g^{\ax\db}\, g^{\cx\bb}\sqrt{g} d^6\xi \ .
\ee
The intersection numbers of $Y_3$ are given by
\be\label{intersection}
\kappa_{abc}=\int \omega^a \wedge \omega^b 
\wedge \omega^c\ .
\ee
Both, the metric and the volume, can be expressed
in terms of the moduli $\vv^a$ and the intersection
numbers 
\bea\label{VGdef}
V&=& \frac{1}{6}\, \kappa_{abc} \vv^a \vv^b \vv^c\ ,\nonumber\\
G_{ab} &=& - V^{-1} \kappa_{abc} \vv^c + 
\frac{1}{4}\,  V^{-2}
 \kappa_{acd}  \vv^c \vv^d \kappa_{bef} \vv^e \vv^f
\ .
\eea

\subsection{The Lagrangian}
The next step is the construction of the dimensionally
reduced type IIB low energy effective theory.
We follow the procedure outlined in
refs.~\cite{BBO} and reduce the 10-dimensional
action (\ref{Sten}) imposing the self-duality
condition on the field equations by hand.
The resulting field equations 
in $d=4$
can be shown to be the Euler-Lagrange equations
of a 4-dimensional action.
The technical details of the reduction procedure 
are deferred to
the appendix and we only give the final result
\bea\label{Sfour}
\e^{-1} \cL &=&  - \half\, \cR 
-\frac{1}{4}e^{2\D}V(v)\, {(\partial_\mu l)}^2
   -(\partial_\mu \phi_4)^2
-\frac{1}{6}\, e^{-3{\D}}V^{\frac12}(v)\, H^{I}_{\mu\nu\rho}M_{IJ}H^{J\mu\nu\rho}\nonumber\\
&& 
 - G_{ab}(v)\,\partial_{\mu}v^a\partial^{\mu}v^b
 - e^{{\D}}V^{\frac12}(v)\, G_{ab}(v)\, 
\partial_{\mu} b^{Ia}M_{IJ}\partial^{\mu}b^{Jb}
\\
&&-\frac{2}{3}e^{-2{\D}}V(v)
\,G_{ab}(v)\,(F^a_{\mu\nu\rho} + 
{\epsilon}_{IJ} H^{J}_{\mu\nu\rho}b^{Ia})
(F^{b\mu\nu\rho} + 
{\epsilon}_{IJ} H^{J\mu\nu\rho}b^{Ib})
\nonumber\\
&& -\frac{i}{12\e}\, {\epsilon}^{\mu\nu\rho\sigma}
\kappa_{abc}{\epsilon}_{IJ}\, 
(F^a_{\mu\nu\rho}b^{Ib}\partial_{\sigma}b^{Jc}
-\frac{2}{3}\,{\epsilon}_{KL}
H^{I}_{\mu\nu\rho}b^{Ja}b^{Kb}\partial_{\sigma}b^{Lc}
)\ ,\nonumber
\eea
where we redefined the field variables
compared to the expansion parameters of 
eqs.~(\ref{harmonic}) according to
\be
D^a_{\mu \nu} \equiv \hat D^a_{\mu \nu} + \half\, \epsilon_{IJ}
B^I_{\mu \nu} b^{Ja}\ , \qquad
F^a_{\mu\nu\rho}\equiv  \partial_{[\mu}D^a_{\nu\rho]}\
,\qquad b^{Ia}\equiv \frac23\, \hat b^{Ia} .
\ee
Furthermore, the kinetic terms for
the $(1,1)$-moduli $\hat v^a$ and the dilaton
decoupled due to the definition
of the four-dimensional dilaton \cite{BCF} 
\be\label{defD}
e^{-2\D} = V(v)\, e^{-2\phi}\ ,
\ee
and `rotated' moduli fields 
\be\label{defv}
v^a = \vv^a e^{\phi/2}\ .
\ee
Note that (\ref{VGdef}) and  (\ref{Sfour}) are
only valid in 
the limit where the 10-dimensional theory
is a good approximation, i.e.\
when $V$ is large.
This limit is termed the `large volume limit'
but  generically subleading $(\alpha')$
corrections
are also important. These are briefly discussed 
in section~4.3.

As we already discussed 
an antisymmetric tensor in four spacetime dimensions
is dual to a scalar
and so we can dualize the 
$h_{(1,1)}+2$ antisymmetric tensor fields
$D^a_{\mu \nu},
B^I_{\mu \nu}$ in the action (\ref{Sfour}).
This is done most conveniently by adding $h_{(1,1)}+2$
Lagrange multipliers $g_a, h_I$ to the Lagrangian
\be\label{LM}
\cL' = \cL + i{\epsilon}^{\mu\nu\rho\sigma}
H^I_{\mu\nu\rho} \partial_\sigma h_I
+ i{\epsilon}^{\mu\nu\rho\sigma}
F^a_{\mu\nu\rho} \partial_\sigma g_a \ .
\ee
The (algebraic) equations of motion
of $\cL'$ for $F^a_{\mu \nu\rho}$ and
$H^I_{\mu \nu\rho}$ can then be used to eliminate
these fields in favour of the scalars $g_a$ and $h_I$.
The resulting Lagrangian reads 
\bea\label{SIIBdual}
\e^{-1} \cL' &=&  - \half \cR 
-\frac{1}{4}e^{2\D}V{(\partial_\mu l)}^2
   -{(\partial_\mu \D)}^2
- G_{ab}\partial_{\mu} b^{1a}\partial^{\mu}b^{1b}\nonumber\\
&& 
 - G_{ab}\partial_{\mu}v^a\partial^{\mu}v^b
 - e^{2\D} V G_{ab} \big(\partial_{\mu} b^{2a}- l\partial_{\mu} b^{1a})
   (\partial^{\mu} b^{2b}-l\partial^{\mu} b^{1b})\nonumber\\
&&
 -\frac{e^{2\D}}{16V}\, {(G^{-1})}^{ab}
  \left(\partial_{\mu}g_a - \half\kappa_{acd}\epsilon_{IJ}
  b^{Ic}{\partial_{\mu}}b^{Jd}\right)
  \left(\partial^{\mu}g_b - \half\kappa_{bef}\epsilon_{KL}
  b^{Ke}{\partial^{\mu}}b^{Lf}\right)\nonumber\\
&&
 -\frac{1}{4}e^{2\D}V^{-1}\left(
\partial_{\mu}h_2
- b^{1a}(\partial_{\mu}g_a
  -\frac{1}{6}\kappa_{abc}\epsilon_{IJ}
  b^{Ib}{\partial_{\mu}}b^{Jc})\right)^2\\
&&
 -\frac{1}{4}e^{4\D}\left(\partial_{\mu}h_1 + l\partial_{\mu}h_2
-
(lb^{1a}-b^{2a})\, (\partial_{\mu}g_a 
 -\frac{1}{6}\kappa_{abc}\epsilon_{IJ}
  b^{Ib}{\partial_{\mu}}b^{Jc})
    \right)^2\ ,\nonumber
\eea
where in addition
the 
$\D$-dependence is displayed more explicitly.

\subsection{Symmetry Properties}
The symmetries of the 10-dimensional theory
discussed in section~2 are also manifest in $d=4$.
The action (\ref{Sfour}) is invariant 
under $h_{(1,1)}+2$ gauge transformations
(with parameters $\Sigma^a_{\nu},\Omega^I_{\nu}$)
of the antisymmetric tensors
\be\label{PQ1}
\delta D^a_{\mu \nu}=\partial_{[\mu}\Sigma^a_{\nu]}\ ,\qquad
\delta B^I_{\mu \nu}= \partial_{[\mu}\Omega^I_{\nu]}\ .
\ee
In addition we have $2\times h_{(1,1)}$
continuous Peccei-Quinn (PQ) symmetries
(with parameters $c^{Ia}$)
acting on  the scalar fields $b^{Ia}$
and the antisymmetric tensors $ D^a_{\mu \nu}$
\be\label{PQ2}
\delta b^{Ia} = c^{Ia}\ , \qquad
\delta D^a_{\mu \nu} = - \epsilon_{IJ}\, c^{Ia} B^J_{\mu \nu}\ .
\ee
These symmetries are `inherited' from the 
ten-dimensional symmetries (\ref{Dsymm}),
(\ref{symmten}).
Note that in the dual basis 
the gauge transformations (\ref{PQ1}) manifest
themselves as $h_{(1,1)}+2$
additional continuous PQ symmetries
(with parameters $ \tilde c_{a},\hat c_I$)
\be\label{PQ3}
\delta g_{a} = \tilde c_{a}\ , \qquad
\delta h_I = \hat c_I\ ,
\ee
while (\ref{PQ2}) transmogrifies into 
\be\label{PQ4}
\delta b^{Ia} = c^{Ia}\ , \quad
\delta g_{a} = \half\, 
\kappa_{abc} \epsilon_{IJ} c^{Ib} b^{Jc}  \ , \quad
\delta h_I = -\epsilon_{IJ}(c^{Ja}g_a 
+\frac16\, \kappa_{abc}
\epsilon_{KL} b^{Ja} c^{Kb} b^{Lc}) 
\ .
\ee

One of the distinct features of the 
type IIB theory in $d=10$ is its 
\SLtwo\ invariance (\ref{SL1})-(\ref{SL3}). 
This symmetry is 
of importance since it includes a
strong-weak coupling duality as one of its
generators.
Thus it is of interest to study the
fate of this symmetry in the $d=4$ action.
Since eq.~(\ref{Sfour}) was obtained 
as a straight dimensional reduction 
of the action (\ref{Sten}) the \SLtwo\
will also be manifest in (\ref{Sfour})
but due to the field redefinitions
(\ref{defD}), (\ref{defv}) it also acts
on the Calabi-Yau moduli $v^a$
and altogether becomes more involved.
More precisely, using (\ref{SL1})-(\ref{SL3}),
(\ref{defD}) and  (\ref{defv}),
one finds
\be\label{SLfour1}
v^a \mapsto v^a |c\lambda +d|\ , \qquad
e^{-2\D} \mapsto {e^{-2\D} \over |c\lambda +d|}
\ ,\qquad
l \mapsto {ac|\lambda|^2 +(bd+ad)\, l+bd
\over |c\lambda +d|^2}\
 ,
\ee
where $\lambda = l + i\, V^{-\frac12}\, e^{-\D}$.
The antisymmetric tensors $B^I_{\mu\nu}$
and the scalars $b^{Ia}$ inherit the transformation
law (\ref{SL1}) and obey
\bea\label{SLfour2}
B^I_{\mu\nu} &\mapsto& 
 \Lambda^I_J B^J_{\mu\nu}\ ,\nonumber\\
b^{Ia} &\mapsto& 
 \Lambda^I_J b^{Ja}\ ,
\eea
while the $D^a_{\mu\nu}$ remain invariant.
{}From (\ref{LM}) one 
infers that the dual scalars $g_a$ 
are  invariant
whereas the $h_I$ transform with the inverse matrix
\be\label{SLfour3}
h_I\mapsto  \Lambda^{-1J}_I h_J\ .
\ee
It can be explicitly checked that 
the Lagrangian (\ref{Sfour}) is invariant 
under the transformations
specified in (\ref{SLfour1}) and (\ref{SLfour2})
and thus at the string tree level and in 
the large volume limit the \SLtwo\
is present in $d=4$ \cite{berkovits}.

\section{The Map between IIA and IIB}
\subsection{Preliminaries}
Type IIA string theory compactified
on Calabi-Yau threefolds features
$h_{(1,1)}$ vector multiplets,
$h_{(1,2)}$ hypermultiplets
and one tensor multiplet in its low energy 
spectrum. The tensor multiplet can be dualized
to an additional hypermultiplet.
A type IIB compactification on
$Y_3$ is equivalent
to a type IIA compactification on the mirror
manifold $\tilde Y_3$ once the appropriate
(worldsheet) 
quantum corrections are taken into account.
This equivalence is a property of the 
underlying conformal field theory \cite{mirror}
but has
also been demonstrated for the geometrical
compactifications on Calabi-Yau manifolds
\cite{CDGP}.
On the mirror manifold  the Hodge numbers 
are reversed and one has
\be
h_{(1,1)}(Y_3) =  h_{(1,2)} (\tilde Y_3)\ ,
\qquad 
h_{(1,2)}(Y_3) =  h_{(1,1)} (\tilde Y_3)\ .
\ee
The low energy spectrum of the two theories
is most easily compared
in the dual basis where the  
$h_{(1,1)}+1$ hypermultiplets of
IIB are identified with the 
$h_{(1,2)}+1$ hypermultiplets of
type IIA.
The purpose of this section 
is to explicitly display
the map between the field variables
of type IIB  and type IIA. 
For  $h_{(1,1)}(Y_3)=1$ this map was given 
in ref.~\cite{BC}. 

Let us start by recalling the low energy effective
action of type IIA.
The universal hypermultiplet of IIA which is
the dual of the tensor multiplet
contains the dilaton $\DA$, 
the dual $\Da$ of an
antisymmetric tensor and two scalars
$\zeta^0, \tilde\zeta^0$ from the R-R sector.
The scalars of the remaining
$h_{(1,2)}$ hypermultiplets are denoted by
$z^a, \bar z^{\bar a}, \zeta^a, \tilde\zeta^a$
where $a=1,\ldots,h_{(1,2)}$.\footnote{%
Here we use a slightly inconsistent notation
since we reserved the index $a$ to label
$(1,1)$-forms. However, we are anticipating
the mirror map between type IIA and type IIB
which identifies the $(1,1)$-forms on 
$Y_3$ with the $(1,2)$-forms on the mirror 
manifold $\tilde Y_3$ and for that reason we 
use the same indices already at this stage.}
The $z^a, \bar z^{\bar a}$ arise from the
NS-NS sector while the $\zeta^a, \tilde\zeta^a$
come from the R-R sector.
The tree level Lagrangian for these hypermultiplets
is known to be \cite{ferrara,BCF}
\bea\label{LIIA}
\e^{-1}\cL & = & -\frac12\, \cR - \,(\partial_\mu \DA)^2
- G_{a b}(z,\bar z)\,\partial^\mu z^a \partial_\mu \bar z^{\bar b}\nonumber\\
&&-\frac14\, e^{4\DA} 
\left(\partial_\mu \Da + \zeta^i \partial_\mu \tilde\zeta_i 
-\tilde\zeta^i \partial_\mu \zeta_i \right)^2
+\frac12\, e^{2\DA}R_{ij}(z,\bar z)\,
\partial^\mu \zeta^i \partial_\mu \zeta^j\\ 
&&+\frac12\, e^{2\DA} R^{-1ij}(z,\bar z)\, \left( I_{ik}(z,\bar z)\,
\partial^\mu \zeta^k +\partial^\mu
  \tilde\zeta_i \right)
\left( I_{jl}(z,\bar z)\,
\partial_\mu \zeta^l +\partial_\mu
  \tilde\zeta_j \right) \ ,\nonumber
\eea
where $i,j,k,l=0,\ldots, h_{(1,2)}$.
The scalar manifold spanned by the 
$4\times(h_{(1,2)}+1)$ scalars is constrained
by $N=2$ supergravity to be a quaternionic
manifold \cite{bw} and this has been explicitly
verified for (\ref{LIIA}) in ref.~\cite{ferrara}.
However, this scalar  manifold is not
the most general quaternionic manifold
but at the string tree level further constrained
by the c-map \cite{cfg}.
That is, the submanifold spanned by the
deformations of
the complex structure 
$z^a$ has to be a special K\"ahler manifold 
\cite{wp}.\footnote{%
The reason is 
that in type IIB vacua the 
$(1,2)$-forms reside in vector multiplets 
and the geometry of their scalars is constrained
to be special K\"ahler.}
More precisely, the metric $G_{a b}(z,\bar z)$
is K\"ahler and furthermore determined
by a holomorphic prepotential 
$F(X)$ which is a homogenous function 
of degree two ($F(\lambda X)=\lambda^2 F(X)$).
Specifically, one has
\be\label{GK}
G_{a b}\ =\ {\partial\over \partial z^a}
{\partial\over \partial\bar z^{\bar b}}\  K\ ,
\ee
where
\bea\label{homcoord}
K&=& -\log[X^i \bar F_i(\bar X) + \bar X^i F_i(X)]\ ,
\nonumber \\
 F_i(X) &\equiv& {\partial F\over \partial X^i}\ ,
\qquad
z^i=\frac{X^i}{X^0}\, \quad (z^0=1)\ .
\eea
%
Furthermore, the matrices 
$R$ and $I$ in (\ref{LIIA})
are also determined by the prepotential
according to 
\bea\label{RIdef}
 R_{ij} &=& \ \mbox{Re}\,\cN_{ij}\ ,\qquad 
  I_{ij} \ = \ \mbox{Im}\,\cN_{ij}\ ,\nonumber \\
  \cN_{ij} &=& \quart  \bar{F}_{ij} - 
               \frac{ (Nz)_i (zN)_j }{(zNz)}\ ,
\eea
where 
\bea\label{FNdef}
  F_{ij} &=& {\partial^2 F\over \partial X^i \partial X^j}\ ,
  \qquad
  N_{ij} = \quart ( F_{ij} + \bar{F}_{ij} )\ ,
\nonumber\\
 (Nz)_i &=& N_{ij} z^j\ , \qquad 
(zNz) = z^i N_{ij} z^j\ .
\eea

\subsection{The Map}
The geometrical compactification
of the effective low energy
IIB supergravity from $d=10$ to $d=4$ 
is only valid
in the large volume limit of  $Y_3$.
In this limit 
the Calabi-Yau volume $V$ is a cubic function
of the $(1,1)$ moduli $v^a$ and 
the intersection numbers
$\kappa_{abc}$ are constant.
In order to display the map between IIA 
and IIB one has to take a similar limit
on the IIA side.
This limit -- termed the `large complex structure
limit' -- has been studied in the 
literature \cite{CDGP,morri}.
For us the important point is that such a 
limit exists 
and in this limit  the prepotential 
of the complex structure moduli is given by
\be\label{Fdef}
F\ =\ \frac{i}{3!} \,
\kappa_{abc}\ {X^a X^b X^c\over X^0}
\ \equiv\  (X^0)^2\, f(z)\ ,
\qquad
f(z) = \frac{i}{3!}\,\kappa_{abc} z^a z^b z^c 
\ .
\ee
Using (\ref{GK}), (\ref{homcoord}),
(\ref{Fdef})
and $z^a = x^a + i y^a$
one obtains the K\"ahler potential and 
the metric 
\bea\label{IIAmet}
  K&=& - \ln \big( \kappa yyy \big)\ , \\  
  G_{a b}(y) &=&
  -\frac{3}{2} \left( \frac{(\kappa y)_{ab}}{(\kappa yyy)}
  -\frac{3}{2} \frac{(\kappa yy)_a (\kappa yy)_b} 
                    {(\kappa yyy)^2} \right)\ , \nonumber
\eea
where we  abbreviated
\be
  (\kappa yyy) = \kappa_{abc} y^ay^by^c\ , \quad
  (\kappa yy)_a = \kappa_{abc} y^by^c\ , \quad
  (\kappa y)_{ab} = \kappa_{abc} y^c\ .
\ee
The matrices defined in (\ref{RIdef})
simplify considerably and are given by  
\bea\label{mat}
  I&=& \frac14 \left(
  \ba{cc} - \frac{1}{3} (\kappa xxx) &
      \frac{1}{2} (\kappa xx)_b \\
      \frac{1}{2} (\kappa xx)_a  &
      -  (\kappa x)_{ab}
  \ea \right)\ , \nonumber\\
  R&=& - \frac{1}{24}(\kappa yyy) \left(
  \ba{cc} 1
          - 4G_{a {b}} x^a x^b &
          4 G_{a {b}} x^b \\
          4 G_{a {b}} x^a &
          - 4G_{a {b}}
  \ea \right)\ , \\
  R^{-1} &=&  - 24 (\kappa yyy)^{-1} \left( 
  \ba{cc} 1 &
          x^b \\
          x^a &
      - \quart G^{-1 a {b}} + x^a x^b
  \ea \right)\ . \nonumber 
\eea
Inserting (\ref{mat}) into  (\ref{LIIA}) 
yields
\bea\label{SIIAL}
  \e^{-1} \cL &=&
  - \half \cR - (\partial_{\mu} \DA)^2 
  - G_{ab} \partial_{\mu} y^a \partial^{\mu} y^b
  - G_{ab} \partial_{\mu} x^a  \partial^{\mu} x^b
  \nonumber \\ &&
  - \frac{1}{8} e^{2\DA} V (\partial_{\mu} \zeta_0)^2
  - \frac{1}{2} e^{2\DA} V G_{ab} 
  \big( x^a \partial_{\mu} \zeta_0 - \partial_{\mu} \zeta^a \big)
  \big( x^b \partial^{\mu} \zeta_0 - 
\partial^{\mu} \zeta^b \big)
  \nonumber \\ &&
  - \half  e^{2\DA} V^{-1} G^{-1ab}
  \Big( \partial_{\mu} \tilde \zeta_a 
        + \frac{1}{8} (\kappa xx)_a \partial_{\mu} \zeta_0
        - \frac{1}{4} (\kappa x)_{ac} \partial_{\mu} \zeta^c \Big)\times
  \nonumber \\ && \phantom{- \half  e^{2\DA} V^{-1} G^{-1ab}}
  \Big( \partial^{\mu} \tilde \zeta_b 
        + \frac{1}{8} (\kappa xx)_b \partial^{\mu} \zeta_0
        - \frac{1}{4} (\kappa x)_{bd} \partial^{\mu} \zeta^d \Big)
  \nonumber \\ &&
  -2  e^{2\DA}V^{-1}
  \Big( \partial_{\mu} \tilde \zeta^0 +  x^a \partial_{\mu} \tilde \zeta_a
        + \frac{1}{24} (\kappa xxx) \partial_{\mu} \zeta_0
        - \frac{1}{8} (\kappa xx)_a \partial_{\mu} \zeta^a  \Big)^2
  \nonumber \\ &&     
  - \quart  e^{4\DA} \big( \partial_\mu \Da + \zeta^i \partial_\mu \tilde\zeta_i 
                            -\tilde\zeta^i \partial_\mu \zeta_i \big)^2\ ,
\eea
where in anticipation of the map to type IIB we 
use 
$V=\frac{1}{6} (\kappa yyy)$ 
(although this does not correspond
to the volume of type  IIA). 

The map of the field variables is obtained by 
comparing the two Lagrangians
(\ref{SIIBdual}) and (\ref{SIIAL}).
First of all one learns that the dilatons 
of the the two theories have to be identified
($\DA=\D$) which is in accord with
the fact that mirror symmetry is already valid in 
string perturbation theory.
Furthermore, one is led to the straightforward 
identification
\be\label{maps}
   y^a = v^a\ , \qquad
   x^a = b^{1a}\ .
\ee
The remaining terms can be systematically
compared due to their specific dependence on
$y^a$ and $\DA$. One obtains  
\bea\label{maps2}
  && \partial_{\mu} \zeta_0 = \pm \sqrt{2} \partial_{\mu} l\ , 
  \nonumber  \\ &&
  b^{1a} \partial_{\mu} \zeta_0 - \partial_{\mu} \zeta^a 
  = \pm \sqrt{2} \big( l\partial_{\mu}b^{1a}-\partial_{\mu}b^{2a} \big)\ , 
  \nonumber  \\ &&
  \partial_{\mu} \tilde \zeta^a 
  + \frac{1}{8} (\kappa b^1b^1)_a \partial_{\mu} \zeta_0
  - \frac{1}{4} (\kappa b^1)_{ab} \partial_{\mu} \zeta^b
  \nonumber  \\ &&
  = \pm \frac{\sqrt{2}}{4}
  \Big( \partial_{\mu} g_a - \half \kappa_{abc}    
  (b^{1b} \partial_{\mu}b^{2c} - b^{2b} \partial_{\mu}b^{1c}) \Big)\ , 
  \nonumber \\ &&
  \partial_{\mu} \tilde \zeta^0 +  b^{1a} \partial_{\mu} \tilde \zeta_a
  + \frac{1}{24} (\kappa b^1b^1b^1) \partial_{\mu} \zeta_0
  - \frac{1}{8} (\kappa b^1b^1)_a \partial_{\mu} \zeta^a
  \nonumber \\ &&
  = \pm \frac{\sqrt{2}}{4}
  \Big( \partial_{\mu} h_2 - b^{1a}\partial_{\mu} g_a 
  + \frac{1}{6} \kappa_{abc}b^{1a} 
  (b^{1b} \partial_{\mu}b^{2c} - b^{2b} \partial_{\mu}b^{1c}) \Big)\ ,
  \\ &&
  \partial_\mu \Da - \tilde\zeta^i \partial_\mu \zeta_i 
  + \zeta^i \partial_\mu \tilde\zeta_i
  \nonumber \\ &&
  = \pm \Big( \partial_{\mu} h_1 + l \partial_{\mu} h_2
  + b^{2a} \partial_{\mu} g_a - l b^{1a} \partial_{\mu} g_a
  - \frac{1}{6} \kappa_{abc}b^{2a} 
    (b^{1b} \partial_{\mu}b^{2c} - b^{2b} \partial_{\mu}b^{1c})  
  \nonumber \\ &&\phantom{\pm \Big(}     
    + \frac{1}{6} l \kappa_{abc}b^{1a} 
    (b^{1b} \partial_{\mu}b^{2c} - b^{2b} 
\partial_{\mu}b^{1c}) \Big)\ .\nonumber  
\eea

\vskip20pt
\noindent
This system of equations is solved by
\bea\label{map}
  \zeta^0 &=& \sqrt{2} l\ ,
  \nonumber \\
  \zeta^a &=& \sqrt{2} (l b^{1a} - b^{2a})\ ,
  \nonumber \\
  \tilde \zeta_a &=& - \frac{\sqrt{2}}{4} g_a 
                   + \frac{\sqrt{2}}{8} l (\kappa b^1b^1)_a
                   - \frac{\sqrt{2}}{8} (\kappa b^1b^2)_a\ ,
   \\
  \tilde \zeta_0 &=& \frac{\sqrt{2}}{4} h_2
                   - \frac{\sqrt{2}}{24} l (\kappa b^1b^1b^1)
                   + \frac{\sqrt{2}}{24} (\kappa b^1b^1b^2)\ ,
  \nonumber \\
  \Da &=& h_1 + \half lh_2 +\half b^{2a} g_a - \half lb^{1a} g_a
        - \frac{1}{12} (\kappa b^1b^2b^2)
        + \frac{1}{12} l (\kappa b^1b^1b^2) \ ,
\nonumber
\eea
which gives the desired relation between
the type IIA and type IIB field variables.
It is somewhat surprising that this relation
is so involved. 
Remarkably (\ref{map}) can be inverted
\bea\label{mapinv}
  l &=& \frac{1}{\sqrt{2}} \zeta_0\ ,
  \nonumber \\
  b^{2a} &=& \frac{1}{\sqrt{2}} (x^a \zeta_0 - \zeta^a)\ ,
  \nonumber \\
  g_a &=& - \frac{4}{\sqrt{2}} \tilde \zeta_a 
          + \frac{1}{2\sqrt{2}} (\kappa x \zeta)_a\ ,
   \\
  h_2 &=& \frac{4}{\sqrt{2}} \tilde \zeta_0
          + \frac{1}{6\sqrt{2}} (\kappa xx \zeta)\ ,
  \nonumber \\
  h_1 &=& \Da- \zeta^i \tilde \zeta_i 
          + \frac{1}{6} (\kappa x \zeta \zeta ) 
          - \frac{1}{12} \zeta_0 (\kappa  x x \zeta )\ .\nonumber
\eea
\vskip 20pt

\subsection{Worldsheet Instanton Corrections}
After having established the map between 
the type IIA and type IIB variables
in the large volume/large complex structure 
limit we can discuss the corrections away from 
this limit.
Mirror symmetry assures the equivalence of 
IIA and IIB compactified on mirror manifolds
and it becomes a matter of convenience in what
variables one describes the action.
The $\alpha'$ corrections to the large volume
limit are most easily discussed in terms
of type IIA variables and
the holomorphic prepotential $f(z)$.
In the large volume limit
$f(z)$ is cubic  (cf.\ (\ref{Fdef})) 
while sub-leading ($\alpha'$) corrections
are known to be \cite{CDGP,CDGPII,HM}
\be\label{fexact}
f(z) = \frac{i}{3!} \,
\kappa_{abc} {z^a z^b z^c}
+ c \chi\zeta(3) + \sum_{d_a} n_{d_a} 
Li_3(e^{-2\pi d_a z^a})
\ .
\ee
Here $c$ is a normalization factor,
$\chi=2(h_{(1,1)}-h_{(1,2)})$ is the 
Euler number, 
$d_a$ is an $h_{(1,1)}$-dimensional summation index,
$n_{d_a}$ are  $h_{(1,1)}$ model dependent
integer coefficients and
\be
Li_3(x)\ =\ \sum_{j=1}^{\infty}\, {x^j\over j^3}\quad .
\ee
In order to obtain the type IIB
effective action (in terms of type IIA
variables) one inserts
(\ref{fexact}) into  (\ref{LIIA}) using 
(\ref{GK})--(\ref{FNdef}).
The same action can be expressed in terms
of type IIB variables by using the transformations
(\ref{maps}) and (\ref{map}).
This last step is necessary in order to discuss
the fate of the \SLtwo\ symmetry 
away from the large volume limit.
It is  this aspect we now turn to.

\subsection{Symmetry Properties}
Let us first identify the PQ symmetries of 
type IIB in the type IIA Lagrangian (\ref{LIIA}).
It is invariant under  
$2\times h_{(1,2)} + 3$ continuous PQ symmetries 
(with parameters $\gamma_i,\tilde\gamma_i,\alpha$)
which act on the fields as follows
\be\label{PQIIA}
\delta \zeta_i = \gamma_i\  , \qquad
\delta \tilde\zeta_i = \tilde \gamma_i\
,\qquad 
\delta \Da = \alpha + \tilde\gamma_i \zeta^i - \gamma_i \tilde \zeta^i\ .
\ee
The $2\times (h_{(1,2)} + 1)$ symmetries of the
first two terms are due to the fact that
$\zeta_i$ and $\tilde\zeta_i$ arise in the R-R sector
while the last symmetry is the PQ
symmetry which any scalar dual to an 
antisymmetric tensor has. 
In the large complex structure limit 
the Lagrangian (\ref{SIIAL}) is invariant under
 $h_{(1,2)}$ additional
continuous PQ symmetries 
which act  on  $x^{a}$ but also on 
$\zeta^i$ and $\tilde \zeta^i$ \cite{dewit}\footnote{%
Note that this symmetry is only given in
its infinitesimal form.}
\bea\label{PQp}
    \delta x^a &=& \hat\gamma^a\ ,  
    \quad
    \delta y^a = 0\ , \quad
 \delta\zeta^0 = 0\ ,
\\
 \delta\tilde \zeta^0 &=& -  \hat\gamma_a \tilde \zeta^a\ , \quad
    \delta\zeta^a = \hat\gamma^a \zeta^0\ ,
    \quad 
    \delta\tilde \zeta^a = \quart 
\kappa_{abc} \hat\gamma^b \zeta^c \ . \nonumber
\eea
The total number of $(3\times h_{(1,2)} + 3)$
PQ symmetries in the type IIA theory
correspond to
the $(3\times h_{(1,2)} + 2)$ PQ symmetries
of type IIB displayed in
(\ref{PQ3}), (\ref{PQ4}) together with
one of the generators of the \SLtwo.  
More precisely, the \SLtwo\ 
of eqs.~(\ref{SL1}) can be 
generated by
\be\label{SLg}
\lambda \mapsto \lambda + 1 \ , \qquad 
\lambda \mapsto -{1\over \lambda}\ ,
\ee
where the first transformation is nothing but 
the `missing' PQ symmetry.\footnote{%
Since at the tree level the symmetry group is really
$SL(2,{\bf R})$ this PQ symmetry is also continuous.}

Worldsheet instantons break 
(\ref{PQp}) to a discrete subgroup.
It is a generic feature of perturbative
string theory that continuous PQ symmetries
of scalars arising in the NS-NS sector
are broken to a discrete subgroup
by strong coupling 
effects on the worldsheet.\footnote{
An exception to this rule are the scalars 
which are dual to antisymmetric tensors
of the NS-NS sector. For example
the PQ symmetry of $\Da$ in 
eq.~(\ref{PQIIA}) is preserved in 
string perturbation theory.} 
On the other hand PQ symmetries
of fields in the R-R sector 
are protected and survive the perturbative
expansion of string theory.
This discussion also applies for the \SLtwo; 
the first transformation of (\ref{SLg})
shifts $l$ of the R-R sector 
and is preserved in perturbation theory. 
The second transformation of (\ref{SLg})
transforms the dilaton
and in section 3.3 we observed
that the entire (\ref{SLg})
is a symmetry in the large volume limit.
In order to discuss the fate of the symmetry
away from this limit we need 
the transformation properties of
the IIA variables in slightly more detail.
Using (\ref{maps}), (\ref{map}) (\ref{mapinv}) and 
(\ref{SLfour1})--(\ref{SLfour3})
one derives for the second generator of 
(\ref{SLg}) the following
transformation law of the type IIA variables
\bea\label{SLmess}
 e^{-2\D} &\mapsto& 
\frac{\sqrt{2}\, e^{-2\D} }{\sqrt{\zeta_0^2+2V^{-1}e^{-2\D}}}\ , 
\nonumber \\
y^a &\mapsto& 
\frac{1}{\sqrt 2}\, 
\sqrt{\zeta_0^2+2V^{-1}e^{-2\D} }\  y^a \ , 
\nonumber \\
x^a &\mapsto& -\frac{1}{\sqrt{2}}\, 
\big( x^a \zeta_0 - \zeta^a \big)\ , 
\nonumber \\
\zeta_0 &\mapsto& - \frac{2\, \zeta_0}{(\zeta_0^2+2V^{-1}e^{-2\D})}\ , 
\nonumber \\
\zeta^a &\mapsto& -\sqrt{2}\,  x^a
+\sqrt{2}\, \frac{\zeta_0(x^a \zeta_0-\zeta^a)}
{(\zeta_0^2+2V^{-1}e^{-2\D})} \ , \\
\tilde \zeta^a &\mapsto& 
\tilde \zeta^a -\frac{1}{4} (\kappa x \zeta)_a
+\frac{1}{8} \zeta_0 (\kappa xx)_a
- \frac{\zeta_0 \kappa_{abc}
(x^b \zeta_0-\zeta^b)(x^c \zeta_0-\zeta^c)}
{8(\zeta_0^2+2V^{-1}e^{-2\D})} \ , 
\nonumber \\ 
\tilde \zeta^0 &\mapsto& 
  - \frac{\sqrt{2}}{4}
    \big( \tilde \phi - \zeta^i \tilde \zeta_i \big)
    +\frac{\sqrt{2}}{48}
    \Big( (\kappa x\zeta\zeta) - \zeta_0 (\kappa xx \zeta) 
           + \zeta_0^2 (\kappa xxx) \Big)
  \nonumber \\ &&
   - \frac{\sqrt{2}}{48}
     \frac{\zeta_0 
           \kappa_{abc}(x^a \zeta_0-\zeta^a)(x^b \zeta_0-\zeta^b)
                       (x^c \zeta_0-\zeta^c)}
          {(\zeta_0^2+2V^{-1}e^{-2\D})}
\ , 
\nonumber \\
\Da &\mapsto& 
\sqrt{2} \Big( 2 \tilde\zeta^0 - x^a \tilde\zeta_a 
                   -\frac{1}{6}(\kappa xx\zeta)
                   + \frac{1}{24} \zeta_0 (\kappa xxx)  \Big)
  \nonumber \\ &&
    + \frac{\zeta_0 \Big( \tilde \phi - \zeta^i \tilde \zeta_i 
                        +2 (x^a \zeta_0-\zeta^a) \tilde\zeta_a
                        +\frac{1}{3}(\kappa x\zeta\zeta) 
                        -\frac{1}{6}\zeta_0 (\kappa xx \zeta) 
                        -\frac{1}{12}\zeta_0^2 (\kappa xxx) \Big)}
           {\sqrt{2}(\zeta_0^2+2V^{-1}e^{-2\D})} 
\ . \nonumber
\eea
Note that the Calabi-Yau moduli 
and the 4-dimensional dilaton
mix in a complicated way
and thus the complex 
$z^a= x^a +i y^a = b^{1a} + i v^a $
transform not into themselves but into a linear
combination involving the $\zeta^i$.
Eqs.\ (\ref{SLmess}) show that
the relatively simple transformations 
(\ref{SLfour1})--(\ref{SLfour3})
become rather involved
in terms of type IIA variables and as a consequence
the \SLtwo\ is hidden in the large
complex structure limit of type IIA. 
Or in other words, the type IIA variables
are not an appropriate basis to
display the \SLtwo\ symmetry.

From (\ref{SLmess}) one also learns
that worldsheet instantons
break the second generator of \SLtwo.
Inserting (\ref{SLmess}) into
(\ref{fexact}) shows
that the prepotential transforms into
a function involving arbitrarily
high powers of the dilaton $e^{-2\D}$
which cannot be cancelled by any other term 
at the string tree level.
Thus strong coupling (non-perturbative) effects
on the string worldsheet 
do not respect the full \SLtwo.
However, it is possible that once non-perturbative
effects in the 4-dimensional space-time are also
taken into account \cite{NP} the \SLtwo\ is restored.
Work along these lines is in progress.

\section{Conclusion}

In this paper we compactified ten-dimensional IIB
supergravity on a Calabi-Yau threefold 
$Y_{3}$ with an arbitrary number 
of harmonic $(1,1)$-forms. 
In the 4-dimensional Lagrangian this 
leads to $h_{(1,1)}$ tensor multiplets coupled
to supergravity and a double-tensor multiplet.
The \SLtwo\ symmetry of 10-dimensional
type IIB theory is also a symmetry of 
the $d=4$ action in the large volume limit
where the 10-dimensional description
is a good approximation.
The \SLtwo\ acts naturally on the 
$d=10$  field variables and therefore 
mixes the 4-dimensional dilaton and the Calabi-Yau
moduli in a non-trivial way.

By mirror symmetry the  type IIB theory 
is equivalent to IIA supergravity 
compactified on the mirror
manifold $\tilde Y_{3}$.
We verified this equivalence 
in the large complex structure limit of type IIA
and displayed  explicitly the relation between the 
two sets of field variables.
Via this map we were able to obtain 
the action of the \SLtwo\ on the type IIA
field variables.
Finally, we noticed that for small Calabi-Yau manifolds 
the \SLtwo\ is broken by
strong coupling effects on the worldsheet
(worldsheet instantons).
 
\begin{appendix}
\section{Appendix}

Throughout this paper we use the following notations and
conventions: the signature of the ten-dimensional 
metric is chosen as $(-++\ldots +)$, capital Latin indices are ten-dimensional indices,
Greek indices from the middle of the alphabet
are four-dimensional space-time indices and the real 
coordinates of the Calabi-Yau threefold 
are denoted by
$y^{\au}; \au=1,\ldots, 6$. We also use a set of complex coordinates
$\xi^{\alpha}, {\bar{\xi}}^{\bar{\alpha}}$ ($\alpha,\bar{\alpha}=1,2,3$), being defined as
$\xi^1=\frac{y^1+iy^2}{\sqrt2},\xi^2=\frac{y^3+iy^4}{\sqrt2},
\xi^3=\frac{y^5+iy^6}{\sqrt2}$
together with their complex conjugates.
The conventions used for Christoffel symbols 
and the various curvature tensors are 
\bea
\Gamma^M_{NP} &=& \half g^{MQ}
(\partial_P g_{NQ}+ \partial_N g_{PQ}- \partial_Q
g_{NP})\ , \nonumber \\
\cR^M_{NPQ}&=&\partial_Q\Gamma^M_{NP}-\partial_P\Gamma^M_{NQ}
+\Gamma^S_{NP}\Gamma^M_{SQ}-\Gamma^S_{NQ}
\Gamma^M_{SP}\ , \\
\cR_{MN} &=& \cR^P_{MPN}\ , \qquad 
\cR=g^{MN}\cR_{MN}\ \nonumber .
\eea

The purpose of this appendix is to supply more 
details on the derivation of the Lagrangian
given in (\ref{Sfour}).
It is constructed by compactifying the $d=10$
effective theory of eq.~(\ref{Sten}) 
on a Calabi-Yau threefold.
First of all the terms which do not involve
the 4-form $\cD_{MNPQ}$ can be treated with standard methods. 
That is, the ten-dimensional metric $g_{MN}$ is decomposed
into a 4-dimensional space-time part $g_{\mu\nu}$
and a 6-dimensional internal Calabi-Yau
metric $g_{\au\bu}$.
This in turn forces a decomposition
of the ten-dimensional Ricci scalar $\cR_{10}$
\be\label{Rdecomp}
\cR_{10}=\cR_4 + g^{\mu \nu}\cR^{\au}_{\mu \au \nu} 
+g^{\au\bu}(\cR^\mu_{\au \mu \bu}+  
\cR^{\cu}_{\au \cu \bu})
 \ .
\ee
The Calabi-Yau metric 
$g_{\au\bu}$ is expanded around a background metric 
$g_{\au\bu}^0$
with small deformations $\delta g_{\au\bu}$
\be\label{CYback}
g_{\au\bu} = g_{\au\bu}^0 + \delta g_{\au\bu}\ .
\ee
In $d=4$
massless scalar fields arise as the inequivalent 
harmonic deformations of the Calabi-Yau metric
\cite{CHSW,CdO}. 
More precisely one expands  
\bea\label{harmonic2}
\delta g_{\ax\bb} &=& 
\vv^a(x)\ \omega^a_{\ax\bb}\ ,\qquad
a=1,\ldots, h_{(1,1)}\ , \nonumber \\
\delta g_{\ax\bx}&=& {z}^A(x)\ \chi^A_{\ax\bx}\ ,
\qquad
A=1,\ldots, h_{(1,2)}\ ,
\eea
where $\omega^a_{\ax\bb}$ are harmonic 
$(1,1)$-forms while $\chi^A_{\ax\bx}$
are related to harmonic $(1,2)$-forms and
$\hat v^a$ and ${z}^A$ are the scalar moduli. 
The ${z}^A$ turn out to be 
members of  vector multiplets
and therefore will be omitted in the following 
discussion. 
Inserting (\ref{CYback}) and
(\ref{harmonic2}) into (\ref{Rdecomp})
and keeping up to quadratic terms in
$\omega^a$ results in \cite{BCF} \footnote{%
There is a slight inconsistency in the derivation 
of eq.~(\ref{Rten}) in ref.\ \cite{BCF} in that not 
all terms contributing at quadratic order are
properly taken into account. 
However, this merely  affects the coefficients
of eq.~(\ref{Rten}). 
We thank M.\ Haack for communicating the correct
formula prior to publication. 
A more detailed derivation of eq.~(\ref{Rten})
will be given
in ref.~\cite{boehm}.}
\bea\label{Rten}
\cR_{10}=\cR_4 -\frac{1}{2}\,
 \partial_{\mu} \vv^a\partial^{\mu} \vv^b\,
\omega^a_{\ax \ab} \omega^{b}_{\bx \bb} 
g^{0\, \ax \bb} g^{0\, \bx \ab}
+ \, \partial_{\mu} \vv^a \partial^{\mu} \vv^b\, 
\omega^a_{\alpha \bar{\alpha}} 
\omega^b_{\beta \bar{\beta}} g^{0\,\alpha \bar{\alpha}} g^{0\,\beta \bar{\beta}} \ .
\eea
The  antisymmetric tensors
$B^I_{MN}$ decompose as
\be\label{Bten}
B^I_{MN} = \left(
\begin{array}{cc}
B^I_{\mu\nu}&0\\
0&B^I_{\au\bu}
\end{array}\right)\ .
\ee
Since there are no harmonic $(2,0)$ forms
on $Y_3$ the $B^I_{\au\bu}$ are expanded only  
in terms of harmonic $(1,1)$-forms
according to eqs.~(\ref{harmonic}).
Finally, the ten-dimensional integration measure
splits according to
\be\label{gten}
\int d^{10}x \sqrt{-g_{10}} = \int d^{4}x \sqrt{-g_4}
\int d^6\xi i\sqrt{g_6}\ ,
\ee
and the integration over the Calabi-Yau threefold
can be performed using 
(\ref{metricdef}) and (\ref{intersection}).

The two remaining terms in eq.~(\ref{Sten})
contain the 4-form $\cD_{MNPQ}$.
As we discussed in section 3.1 
the massless modes in $d=4$
arising from dimensional
reduction of  $\cD_{MNPQ}$ are 
vectors $\cD_{\mu \au\bu\cu}$ 
and tensors $\cD_{\mu\nu\au\bu}$.
Not all of them are independent but related by
the self-duality condition (\ref{selfdual}).
The vectors are of no concern here and so in
the reduction procedure we only focus on the
tensor fields. 
In order to proceed we make an Ansatz 
for the couplings of the tensor fields
where
the individual terms are dictated
by the reduction of the last terms in (\ref{Sten}) 
\bea\label{ansatz}
\cL&=&\sqrt{-g_{10}}
\left(k_1 (F_{\mu \nu \rho \au\bu})^2
+ k_2 {\epsilon}_{IJ}B^I_{\au\bu}H^J_{\mu \nu \rho}
F^{\mu \nu \rho \au\bu}
+ k_3 {\epsilon}_{IJ}{\epsilon}_{KL}B^I_{\au\bu}
H^J_{\mu \nu \rho}
 B^{K\au\bu}H^{L\mu \nu \rho} \right) \nonumber\\
 &&+{\epsilon}^{\mu \nu \rho \sigma}\epsilon^{\au\bu\cu\du\eu\fu} {\epsilon}_{IJ}
 \left(k_4 F_{\mu \nu \rho
\au\bu}B^I_{\cu\du}\partial_{\sigma}B^J_{\eu\fu} 
+ {k_5} {\epsilon}_{KL}
H^I_{\mu \nu
  \rho}B^J_{\au\bu}B^K_{\cu\du}\partial_{\sigma}B^L_{\eu\fu}\right)\ ,
\eea
where
\bea
F_{\mu \nu \rho \au\bu}&\equiv& \frac{1}{30}
(\partial_{\mu}\tilde{\cD}_{\nu \rho \au\bu}
+\partial_{\rho}\tilde{\cD}_{\mu \nu\au\bu}
+\partial_{\nu}\tilde{\cD}_{\rho \mu\au\bu})\ ,
\nonumber\\
\tilde{\cD}_{\mu \nu \au\bu}&\equiv& 
\cD_{\mu \nu\au\bu}
+\frac{3}{4}{\epsilon}_{IJ}B^I_{\mu \nu}B^J_{\au\bu}
\ ,
\eea
and $k_1,\ldots, k_5$ are constants to be determined.
This is done by comparing the equations of
motion of (\ref{ansatz}) 
with the dimensionally reduced 10-dimensional
field equations.
Since we only need to fix a few constants
we simplify the task by doing this comparison
in  the limit $l=0$, $\phi = const.$ and $g_{\mu \nu}=\eta_{\mu \nu}$.
For our purpose the important field equations
turn out to be
\be\label{Feq}
\partial^P G_{MNP}=-\frac{10}{3}iF_{MNPQS}G^{PQS} \ .
\ee
The general definition of $G_{MNP}$ can be found
in \cite{BBO} while here we only 
record its components in this specific limit
\bea
G_{\mu \nu \rho}=e^{\frac{\phi}{2}}(e^{-\phi}H^1_{\mu \nu \rho}+iH^2_{\mu
\nu \rho})\ ,\nonumber\\
G_{\mu \au\bu}
=\frac{1}{3}e^{\frac{\phi}{2}}
(e^{-\phi}\partial_{\mu} B^1_{\au\bu}
+i\partial_{\mu} B^2_{\au\bu})\ ,\\
G_{\au\mu \nu}=G_{\mu \au \nu}
=G_{\mu \nu \au}=G_{\au\bu\cu}
=0 \ .\nonumber
\eea
As a consequence (\ref{Feq})
simplifies to
\bea\label{Feq2}
\partial^{\rho}G_{\rho\mu \nu}
&=& -10i F_{\mu \nu \rho \au\bu}G^{\rho \au\bu}\\
\partial^{\mu}G_{\au\bu\mu}&=&-\frac{10}{3}i(F_{\mu \nu \rho \au\bu}G^{\mu \nu
\rho}+3F_{\mu \au\bu\cu\du}G^{\mu \cu\du})\ .\nonumber
\eea 
The first equation is merely an equation for
 $B^I_{\mu \nu}$ and hence not of immediate
interest for us.
Using the self-duality condition
$F_{\mu \au\bu\cu\du}G^{\mu \cu\du}
=\frac{1}{12\e}\, \epsilon_{\mu \nu \rho \sigma}
\epsilon_{\au\bu\cu\du\eu\fu}
F^{\nu \rho \sigma \eu\fu}G^{\mu \cu\du}$
the second equation in (\ref{Feq2})
yields for $B^1_{\au\bu}$
\bea\label{BOB}
\partial^{\mu}\partial_{\mu} B^1_{\au\bu}
&=&10\, e^{\phi}\, (F_{\mu \nu \rho \au\bu}+\frac{3}{20}
{\epsilon}_{IJ}B^I_{\au\bu}H^J_{\mu \nu \rho})\
H^{2\mu \nu \rho}\\ 
&&+\frac{5}{6\e}\,e^{\phi} \epsilon_{\mu \nu \rho
  \sigma}\epsilon_{\au\bu\cu\du\eu\fu}\, 
(F^{\nu \rho \sigma \eu\fu}
+\frac{3}{20}{\epsilon}_{IJ}B^{I\eu\fu}H^{J\nu \rho
\sigma})\, 
\partial^{\mu}B^{2\cu\du}\ . \nonumber
\eea
This equation has to be compared with the 
equation of motion for $B^1_{\au\bu}$
obtained from (\ref{ansatz})
\bea\label{Fansatz}
\partial^{\mu}\partial_{\mu}B^1_{\au\bu}&=&
-4\, e^{\phi}\,(k_2 F_{\mu \nu \rho \au\bu}
+2k_3 {\epsilon}_{IJ}B^I_{\au\bu}H^J_{\mu \nu \rho})\,  H^{2\mu \nu \rho} \\
&&
-\frac{4}{\e}\, e^{\phi}
{\epsilon}_{\mu \nu \rho \sigma}
\epsilon_{\au\bu\cu\du\eu\fu}
( 2k_4 F^{\mu \nu \rho \cu\du}
-k_5 {\epsilon}_{IJ}B^{I\cu\du}H^{J\mu \nu \rho}
)\, \partial^{\sigma}B^{2\eu\fu} \ . \nonumber
\eea
Comparing (\ref{BOB}) with (\ref{Fansatz})
yields 
\be
k_1 = -\frac{25}{3}\ ,\qquad
k_2 = -\frac{5}{2}\ ,\qquad
k_3 = -\frac{3}{16}\ ,\qquad
k_4 = \frac{5}{48}\ , \qquad 
k_5 =-\frac{1}{32}\ ,
\ee
where the overall normalization is determined by
using the gauge invariance
of eqs.~(\ref{PQ1}) and (\ref{PQ2}).
(The analogous equations for $B^2_{\au\bu}$
are consistent with the same set of coefficients.)
Finally one adds (\ref{ansatz})
to the dimensionally reduced action obtained
from the first three terms in (\ref{Sten}) 
using (\ref{Rten}),
(\ref{Bten}) and (\ref{gten}),
performs a Weyl rescaling 
$g_{\mu \nu}\rightarrow V^{-1}g_{\mu \nu}$
and integrates
over the Calabi-Yau threefold
using the formulae given in section 3.1.
This results in the Lagrangian (\ref{Sfour}).

\end{appendix}

\vskip 1cm

{\large \bf Acknowledgements}

This work is supported in part by
the French-German binational 
program PROCOPE.  
C.H.\ thanks J.L.\ and his group 
for the hospitality in Halle
and H.G. thanks R.\ Grimm
and his group for the hospitality in Marseille.

This work is additionally supported by
GIF -- the German--Israeli
Foundation for Scientific Research.
J.L.\ thanks S.\ Yankielowicz and J.\ Sonnenschein
for the hospitality in Tel Aviv and Y.\ Nir
for the hospitality at the Weizmann Institute.

We thank I.\ Antoniadis, C.\ Bachas, B.\ de Wit,
M.\ Haack, T.\ Mohaupt and  V. Kaplunovsky  
for useful conversations.

\end{document}